\documentclass[12pt]{article}
\usepackage[dvips]{graphicx}
\usepackage{latexsym}
\usepackage{amssymb}
\newcommand{\be}{\begin{equation}}
\newcommand{\bea}{\begin{eqnarray}}
\newcommand{\ee}{\end{equation}}
\newcommand{\eea}{\end{eqnarray}}
\newcommand{\eq}[1]{Eq.~(\ref{#1})}

\newcommand{\bk}{B^0 \to \phi K^{\ast 0}}

\begin{document}

\thispagestyle{empty}
\begin{center}
\hfill FTUV/04$-$1006 \\
\vspace*{2.5cm}
{\Large \bf Right handed currents and FSI phases in $\bk$}
\vspace*{1cm} \\
{ \sc E.\ \'Alvarez$^{a,b}$, L.N.\ Epele$^a$, D.\ G\'omez Dumm$^a$, A.\ Szynkman$^a$}
\vspace*{1.2cm} \\
{\em $^a$ IFLP, CONICET $-$ Depto.\ de F\'{\i}sica,
Universidad Nacional de La Plata, \\
C.C. 67, 1900 La Plata, Argentina \\ $\,$\\
$^b$Departament de F\'{\i}sica Te\`orica, IFIC,
CSIC -- Universitat de Val\`encia \\
Dr.\ Moliner 50, E-46100 Burjassot (Val\`encia), Spain}
\vspace*{3cm}
\begin{abstract}
We consider possible effects of New Physics (NP) on the angular
distributions of the decay $\bk$, showing how these effects depend on the
nature of nonstandard interactions. In a general framework based on
factorization, we show that triple products can be used to probe the
chirality of NP currents. In this analysis we take into account the
presence of non-vanishing strong phases, which is motivated by recent
experimental evidence. It is seen that the observability of right-handed
NP is strongly dependent on the relation between the relative magnitude of
these phases and the ratio of Standard Model and NP scales. As an
application we estimate the expected values of relevant observables in a
particular Left Right Symmetric Model.
\end{abstract}
\end{center}
\vspace*{.5cm}
\hspace*{0.5cm} PACS~: 11.30.Er, 12.60.-i, 13.25.Hw

\newpage

\section{Introduction}

The theoretical and experimental study of B meson physics offers a good
opportunity to get new insight about the origin of CP violation. Taking
into account the rich variety of decays channels, one can look for many
different observables, providing stringent tests for the consistency and
stoutness of the CKM mechanism of CP violation proposed by the Standard
Model (SM). These analyses may definitely unveil the presence of New
Physics (NP) beyond the SM, or provide hints for future searches. In the
last few years, an important goal has been achieved through the
measurements of the time dependent CP asymmetry in $B \to J / \psi K_S$
decays by BABAR~\cite{Babar} and Belle~\cite{Belle}, which have firmly
established the violation of CP in the $B$ system. Within the SM, these
experiments allow to get a ``clean'' measurement of the value of
$\sin(2\beta)$, where $\beta$ is one of the angles of the so-called
unitarity triangle. The present world average value for this quantity
is~\cite{pdg}
\begin{equation}
\sin (2 \beta)_{J/\psi K_S}\ = \ 0.731\ \pm 0.056 \ ,
\label{J}
\end{equation}
which is in agreement with SM expectations. However, the success of the SM
is not clear in the case of CP asymmetries in the channel $B\to\phi K_S$,
where recent measurements of $\sin (2 \beta)$ lead to
\begin{eqnarray}
\sin (2 \beta)_{\phi K_S} & \! = \! & \ \ \, 0.47\ \pm\ 0.34 
\hspace{1cm} {\rm BABAR \mbox{\cite{Babar2}}} \nonumber \\
\sin (2 \beta)_{\phi K_S} & \! = \! & -0.96\ \pm\ 0.51
\hspace{1cm} {\rm Belle \mbox{\cite{Belle2}}} \quad ,
\label{Phi}
\end{eqnarray}
showing an apparent disagreement with the previous result. It is not
unreasonable to expect this discrepancy to be originated by the presence
of NP contributions. Indeed, NP effects are likely to be more important in
the case of $B\to\phi K_S$ than $B\to J/\psi K_S$ since in the SM the
latter is governed by a tree level amplitude, while the former is shown to
occur dominantly through penguin-like processes.

In this work we propose to explore the NP hypothesis by considering a
different channel, namely the decay $\bk$, and its CP conjugate, $\bar B^0
\to \phi \bar K^{\ast 0}$. These decays, which are driven by the same
quark level processes as $B\to\phi K_S$, are found to be attractive for
several reasons. Among the various charmless $B\to VV$ channels, the
neutral and charged $B\to \phi K^\ast$ decays are the first ones that have
been experimentally observed~\cite{CLEO}. In addition, $B\to VV$ processes
offer the possibility of measuring many different observables, taking into
account the angular distributions of the final outgoing states. In fact,
even if $\bk$ is a neutral decay, various asymmetries that are relevant
for the study of CP violation can be determined with no need of either
flavor tagging or time-dependent measurements, hence the experimental
analyses are considerably simplified. Detailed studies of the angular
distributions have been performed recently by BABAR~\cite{Bab04} and
Belle~\cite{Bel04} collaborations, leading to some puzzling results that
deserve significant theoretical interest~\cite{perinola}. Although the
corresponding experimental errors are still relatively large so as to
unveil the presence of NP, the channel is a promising one to encourage the
search. In this sense, it is worth to notice that within the SM the $B^0
\to \phi K^{\ast 0}$ decay amplitude has the feature of being dominated by
penguin contributions carrying approximately a common weak phase.
Therefore, CP-violating observables are expected to be suppressed within
the SM, enhancing the possibility of finding signals of NP.

If the observation of NP effects is confirmed, the various observables to
be measured at $B$ factories should be used to distinguish between
possible extensions of the SM and to determine the corresponding
parameters~\cite{london}. In this article we point out that the angular
distributions of decay products in $B^0 \to \phi K^{\ast 0}$ could provide
not only evidences of physics beyond the SM~\cite{Gir03} but also relevant
information about the {\it nature} of this NP. In particular, using the
factorization approach~\cite{Bau85}, we show that it is possible to
perform a chirality test~\cite{previous} for the structure of nonstandard
effective current-current operators through the measurement of the
so-called triple products~\cite{lonval}. In our analysis we take into
account the presence of non-vanishing strong FSI phases, which is
stimulated by recent experimental results reported by BABAR and
Belle~\cite{Bab04,Bel04}. In this regard, we point out that the expected
order of magnitude of the relevant CP-odd quantities depends on the
interplay between the scale of NP and the size of strong phases. The
proposed chirality test can be used to distinguish between NP effects
arising from right-handed currents (e.g.\ those predicted by left-right
symmetric models) from those which come from SM-like operators, as one
would find for instance in the case of the Minimal Supersymmetric Standard
Model (MSSM), or multi-Higgs models with $SU(2)_L\otimes U(1)_Y$
electroweak gauge symmetry. Finally, in order to illustrate the potential
significance of our analysis, we give an estimation of the size of the
expected effects in the case of a Left-Right Symmetric Model (LRSM).

This paper is organized as follows: in Sec.\ II we write the effective
Hamiltonian in the presence of NP right-handed currents, we identify
different contributions to the amplitude in the helicity basis and we set
the hypotheses to be used in later analysis. In Sec.\ III we explain how
the angular analysis of the decay is used, within factorization, to build
the appropriate observables to test the chirality of NP currents. Some of
these observables are estimated in Sec.\ IV within a particular left-right
symmetric model. Finally, in Sec.\ V we state our conclusions.

\section{Amplitudes and phases in $B\to VV$ decays}

{}From the theoretical point of view, the standard way of dealing with
nonleptonic $B$ decays is based on the effective Hamiltonian approach,
which makes use of the Operator Product Expansion as a fundamental tool.
Within the SM, this program has been developed in detail, including
next-to-leading order calculations~\cite{Buc96}. After integrating out the
degrees of freedom corresponding to all particles with masses above the
$b$ quark scale, the low-energy effective Hamiltonian responsible for
$b\to s$ transitions can be written as
\begin{equation}
H_{\rm eff} = \frac{G_F}{\sqrt{2}} \left[ V_{cb}^\ast V_{cs} \sum_{i=1,2}
C_i\; O_i \ + \ V_{tb}^\ast V_{ts} \sum_{i=3...10,g,\gamma} C_i\; O_i \
\right] . \label{heff}
\end{equation}
where $O_i$ are local current-current quark operators, and $C_i$ are the
corresponding Wilson coefficients evaluated at a renormalization scale
$\mu\approx m_b$. Following the notation of Ref.~\cite{Ali98}, here
$O_{1,2}$ are standard current-current operators, $O_{3\dots 6}$ and
$O_{7\dots 10}$ stand for QCD and EW penguin operators, respectively,
and $O_g$ ($O_\gamma$) denotes the gluonic (photonic) magnetic dipole
operator. Explicit expressions for both the effective operators and the
Wilson coefficients within the SM can be found in several articles and
reports (see e.g.\ Refs.~\cite{Ali98,AliChe}) and we will not repeat them
here. Clearly, this effective Hamiltonian would be in general modified if
one allows for the presence of NP. It could happen, for instance, that NP
effects only show up effectively through SM-like operators, hence the new
contributions would just add some new terms to the coefficients $C_i$.
This is indeed the situation in the context of some popular schemes, such
as the Two-Higgs doublet model and the MSSM. Conversely, it might happen
that NP contributions give rise to new effective operators, to be added to
the $O_i$'s in \eq{heff}. Here we will address the particular case in
which NP manifests effectively through right handed currents, as e.g.\ in
Left-Right (LR) symmetric models~\cite{lrsm}. Within these models one
would find new operators $O'_i$, obtained from the $O_i$'s through the
interchange $1\pm\gamma_5 \to 1\mp\gamma_5$. Moreover, in this case one
should add to $H_{\rm eff}$ four new tree operators, namely
\begin{eqnarray}
O_{11} &=& \frac{m_b}{m_c}\; (\bar s_\alpha \gamma_\mu (1-\gamma_5)\, c^\beta)
\,(\bar c_\beta \gamma^\mu (1+\gamma_5)\, b^\alpha )\ , \\
O_{12} &=& \frac{m_b}{m_c}\; (\bar s_\alpha \gamma_\mu (1-\gamma_5)\, c^\alpha)
\,(\bar c_\beta \gamma^\mu (1+\gamma_5)\, b^\beta )\ ,
\end{eqnarray}
plus $O'_{11}$, $O'_{12}$, which are obtained from the previous two by
interchanging $1\pm\gamma_5 \to 1\mp\gamma_5$. Thus the effective $\Delta
B=1$ Hamiltonian will be given by
\bea H_{\rm
eff} & = & H_{\rm eff}^{\rm (SM)}\ +\ H_{\rm eff}^{\rm (NP,L)}\ +
\ H_{\rm eff}^{\rm (NP,R)}
\label{decomp} \\
 & = &  \sum_{i=1...10,g,\gamma} \tilde C_i\; O_i \ + \sum_{i=1...
12,g,\gamma} \delta \tilde C_i\; O_i \ + \sum_{i=1...12,g,\gamma} \delta
\tilde C'_i\; O'_i\ ,
\label{Heff}
\eea
where the NP coefficients $\delta \tilde C_i$ and $\delta \tilde C'_i$ can
be calculated as before using renormalization group equations in the
extended LR symmetric theory. In \eq{Heff} we have added a tilde to the
Wilson coefficients, which for simplicity have been redefined absorbing
the common factor $G_F/\sqrt2$ and the corresponding CKM matrix elements.
Notice that $\delta \tilde C_i$ correspond to contributions driven by both
SM-like operators ($i=1\dots 10,g,\gamma$) and nonstandard tree operators
($O_{11,12}$).

Given the relevant effective Hamiltonian, the decay amplitude of a $B$
meson into two vector mesons $V_1$ and $V_2$ in a definite helicity state
is written as $A_\lambda = \langle V_1(\lambda) V_2(\lambda)|H_{\rm
eff}|B\rangle$, where $\lambda=0,\pm 1$ is the helicity of both $V_1$ and
$V_2$. According to its Lorentz structure, this amplitude can be
parameterized as~\cite{Kra92}
\begin{eqnarray}
A_{\lambda} = \varepsilon_{1 \mu}^* (\lambda) \; \varepsilon_{2 \nu}^*
(\lambda) \left[\; a\, g^{\mu \nu} + \frac{b}{m_1 m_2}\; p^\mu p^\nu +
\frac{i\; c}{m_1 m_2}\; \epsilon^{\mu \nu \alpha \beta}\, p_{1\alpha}
p_{2\beta} \right]\ ,
\label{hlam}
\end{eqnarray}
where $p$ is the four-momentum of the $B$, while $m_i$, $p_i$ and
$\varepsilon_i$ stand for the masses, momenta and polarization vectors of
the $V_i$ mesons respectively.

In general, the parameters $a$, $b$ and $c$ are complex numbers which
arise from the sum of various interfering contributions. Each of these
contributions will carry in principle different strong (CP-conserving) and
weak (CP-violating) phases, which we will denote as usual by $\delta_i$
and $\phi_i$ respectively. Now, for the sake of clarity, let us assume
that each term in the decomposition of $H_{\rm eff}$ in \eq{decomp}
provides a single dominant contribution (we recall that within the SM this
is the situation in the case of $\bk$). In general, this does not need to
be the case, but the introduction of more terms will not alter our
conclusions qualitatively. Labelling the (SM), (NP,L) and (NP,R)
contributions in \eq{decomp} with 1, 2 and 3 respectively, we write $a$,
$b$ and $c$ as \bea a &=& \left( a_1 e^{i (\tilde\delta_1+\phi_1)} + a_2
e^{i(\tilde \delta_2 + \phi_2)}
+ a_3 e^{i(\tilde \delta_3 + \phi_3)}\right) e^{i\delta_a} \nonumber \\
b &=& \left( b_1 e^{i (\tilde \delta_1+\phi_1)}
+ b_2 e^{i(\tilde \delta_2 + \phi_2)}
+ b_3 e^{i(\tilde \delta_3 + \phi_3)}\right) e^{i\delta_b} \nonumber \\
c &=& \left( c_1 e^{i(\tilde \delta_1+\phi_1)} + c_2 e^{i(\tilde \delta_2
+ \phi_2)} + c_3 e^{i(\tilde \delta_3 + \phi_3)}\right) e^{i\delta_c} \ .
\label{coefficients} \eea Here we have distinguished between strong phases
originated from high energy absorptive contributions, denoted by $\tilde
\delta_i$, and those arising from low energy final state interactions
(FSI), $\delta_{a,b,c}$. It is reasonable to assume that the former come
only through the Wilson coefficients, hence they are common to $a$, $b$
and $c$ for each contribution 1, 2 and 3. The same argument holds also for
the weak phases $\phi_i$. On the contrary, while FSI interaction phases
may be different for $a$, $b$ and $c$, we expect them to appear in each
case as global factors, since low energy FSI should be blind to the
decomposition into standard and nonstandard quark-level operators. In
\eq{coefficients} there is still an overall phase ambiguity, which allows
e.g.\ to fix the SM phase $\tilde \delta_1+ \phi_1=0$ without loss of
generality. Here we have kept all phases as nonvanishing for the sake of
clarity in the forthcoming results. It is important to notice that the
contributions 2 and 3, arising from NP, are expected to be suppressed with
respect to the SM contribution 1. The order of magnitude of the
corresponding suppression factor, say $\xi$, will be given by the ratio
between typical standard and new physics energy scales.

A key point to be remarked is the fact that in \eq{hlam} the Lorentz
tensors associated with coefficients $a$ and $b$ have an opposite behavior
under parity in comparison with the pseudotensor associated with $c$. We
will see below that this can be used to disentangle the different NP
contributions to the effective Hamiltonian, in particular, allowing to
probe the magnitude of the contribution of right-handed currents in
left-right symmetric extensions of the SM. It is crucial to find adequate
observables to achieve this disentanglement, and this is the subject of
the following Section.

\section{Strong phases and observables sensitive to right-handed NP}

Let us consider a decay of a $B$ into two vector mesons, $B\to V_1 V_2$,
followed by the decays $V_1\to P_1 P'_1$ and $V_2\to P_2 P'_2$, where
$P_i$, $P'_i$ are pseudoscalars. The normalized differential angular
distribution can be written as~\cite{Chiang00}
\begin{eqnarray}
\frac{1}{\Gamma_0}\ \frac{d^3\Gamma}{d\cos\theta_1\; d\cos\theta_2\;d\psi}
& = &
\frac{9}{8 \pi {\cal N}} \;\Bigg\{ |A_0|^2 \;\cos^2\theta_1 \;\cos^2\theta_2 \
+ \ \frac{|A_\| |^2}{2} \;\sin^2\theta_1 \;\sin^2\theta_2 \;\cos^2\psi
\nonumber \\
& & \hspace*{-2.7cm}
+ \ \frac{|A_\bot|^2}{2} \; \sin^2\theta_1 \;\sin^2\theta_2 \;\sin^2\psi \
+ \ \frac{{\rm Re}(A_\| A_0^\ast)}{2\sqrt{2}} \;\sin 2\theta_1
\;\sin 2\theta_2 \;\cos\psi \nonumber \\
& & \hspace* {-2.7cm}
- \ \frac{{\rm Im}(A_\bot A_0^\ast)}{2\sqrt{2}} \;\sin 2\theta_1
\;\sin 2\theta_2 \;\sin \psi \
- \ \frac{{\rm Im}(A_\bot A_\|^\ast)}{2} \;\sin^2\theta_1 \;\sin^2\theta_2
\;\sin 2 \psi \Bigg\}
\label{dg}
\end{eqnarray}
where $\theta_1$ ($\theta_2$) is the angle between the three-momentum of
$P_1$ ($P_2$) in the $V_1$ ($V_2$) rest frame and the three-momentum of
$V_1$ ($V_2$) in the $B$ rest frame, $\psi$ is the angle between the
planes defined by the $P_1 P'_1$ and $P_2 P'_2$ three-momenta in the $B$
rest frame, and we have defined ${\cal N}\equiv |A_0^2|+|A_\bot|^2+|A_\|
|^2$. The amplitudes $A_\bot$ and $A_\|$ are related to the helicity
amplitudes introduced in \eq{hlam} by
\begin{equation}
A_{\bot} \; = \; \frac{A_{+1} - A_{-1}}{\sqrt{2}}\;, \qquad
A_{\|} \; = \; \frac{A_{+1} + A_{-1}}{\sqrt{2}}\;,
\label{cb}
\end{equation}
while $A_0$ is common to both bases. In addition, the observables in
\eq{dg} can be written in terms of the previously introduced complex
parameters $a$, $b$ and $c$ as
\begin{eqnarray}
\lefteqn{|A_0|^2\; =\; |x\, a + (x^2-1)\, b|^2} \hspace*{5cm} & &
{\rm Re}(A_\| A_0^\ast)\; =\; -\sqrt{2}\,\left[ x\, |a|^2 + (x^2-1)\,{\rm
Re}(a^\ast b)\right]
\nonumber \\
\lefteqn{|A_\| |^2\; =\; 2\,|a|^2} \hspace*{5cm} & &
{\rm Im}(A_\bot A_0^\ast)\; =\; \sqrt{2\,(x^2-1)} \left[ x\, {\rm Im}
(a c^\ast) + (x^2-1)\,{\rm Im}(b c^\ast)\right] \nonumber \\
\lefteqn{|A_\bot|^2\; =\; 2\,(x^2-1)|c|^2} \hspace*{5cm} & &
{\rm Im}(A_\bot A_\|^\ast)\; =\; 2\sqrt{x^2-1}\; {\rm Im} (c a^\ast) ,
\label{as}
\end{eqnarray}
where $x \equiv (m_B^2 - m^2_1 - m^2_2)/(2 m_1 m_2)$. Now as usual we
denote the amplitudes corresponding to the CP conjugated process by $\bar
A_i$. This comes together with the replacements $a\to\bar a$, $b\to\bar
b$, $c\to -\bar c$ in the parameterization of \eq{hlam}. In addition,
following the notation of Ref.~\cite{Lonsin}, we define the combined
observables
\begin{equation}
\begin{array}{ll}
\Lambda_{ii}\ = \ \frac12 (|A_i|^2 + |\bar A_i|^2)
\ , &
\Sigma_{ii}\ = \ \frac12 (|A_i|^2 - |\bar A_i|^2)\ ,
\\
\rule{0cm}{.7cm}\Lambda_{\bot j} = -\,{\rm Im}(A_\bot A_j^\ast -
\bar A_\bot \bar A_j^\ast)\ , & \Sigma_{\bot j} = -\,{\rm Im}(A_\bot A_j^\ast +
\bar A_\bot \bar A_j^\ast)\ ,
\\
\rule{0cm}{.7cm}\Lambda_{\| 0} = -\,{\rm Re}(A_\| A_0^\ast +
\bar A_\| \bar A_0^\ast)\ , &
\Sigma_{\| 0} = -\,{\rm Re}(A_\| A_0^\ast -
\bar A_\| \bar A_0^\ast)\ ,
\end{array}
\label{obs}
\end{equation}
where $i=0,\|,\bot$, and $j=0,\|$.

Among the kinematic observables in the decay of a $B$ into two vector
mesons, one has the so-called ``triple products'' $\vec{q}_i \cdot
(\vec{\epsilon}_1 \times \vec{\epsilon}_2)$, where $\vec{\epsilon}_1$,
$\vec{\epsilon}_2$ are the polarization vectors of the two vector mesons,
and $\vec{q}_i$ is the momentum of one of them. Since the T-transformation
reverses both momentum and spin, triple products are T-odd quantities.
Regarding the observables defined above, one has that ${\rm Im}(A_\bot
A_{0,\|}^\ast)$ is proportional to the asymmetry
\begin{equation}
A_T\ = \ \frac{\Gamma \, (\vec{q}_i \, \cdot \, (\vec{\epsilon}_1 \,
\times \, \vec{\epsilon}_2) \, > \, 0) \ - \ \Gamma (\vec{q}_i \, \cdot \,
(\vec{\epsilon}_1 \, \times \, \vec{\epsilon}_2) \, < \, 0)}{\Gamma \,
(\vec{q}_i \, \cdot \, (\vec{\epsilon}_1 \, \times \, \vec{\epsilon}_2) \,
> \, 0) \ + \ \Gamma (\vec{q}_i \, \cdot \, (\vec{\epsilon}_1 \, \times \,
\vec{\epsilon}_2) \, < \, 0)} \ ,
\end{equation}
which is also T-odd. Note that this asymmetry may be not null even in the
absence of CP violation, due to the presence of strong
phases~\cite{lonval}. If, as in the case of the combined observables
$\Lambda_{\bot (\| , 0)}$, we define the CP-asymmetry ${\cal A}_T = A_T -
\bar A_T$, where $\bar A_T$ is the triple-product asymmetry corresponding
to the CP-conjugated decay, then ${\cal A}_T$ will be both a T-odd and
CP-odd quantity. Since ${\cal A}_T$ is an observable of CP violation, it
will be also an observable of T violation if CPT invariance is assumed. On
the other hand, the combined observable $\Sigma_{\bot (\| , 0)}$ is
proportional to $A_T + \bar A_T$, therefore it is a T-odd but CP-even
quantity.

Let us focus our attention in the neutral decay $\bk$ (and the
corresponding CP conjugated process, $\bar B^0 \to \phi \bar K^{*0}$) in
the context of a model including left- and right-handed quark currents
arising from new physics, as discussed in Sec.~II. This decay
proceeds through the quark level process $\bar b\to \bar s s \bar s$,
which is mainly driven by penguin operators $O_i$ and $O'_i$, with
$i=3,\dots , 10$ in the notation introduced above. In order to deal with
the hadronic matrix element, we will use the well-known factorization
approximation, in which one neglects the contribution of color-octet quark
transitions. However, we will allow for the presence of nonzero final
state interaction phases. In this way, the standard and nonstandard
contributions to the amplitude can be factorized as
\begin{eqnarray}
A & = &  [\tilde C + \delta \tilde C]\; \langle K^{\ast 0} | \bar b
\gamma_\mu (1 - \gamma_5) s | B^0 \rangle \langle \phi | \bar s \gamma^\mu
s | 0 \rangle\ +
\nonumber \\
& & \rule{0cm}{.5cm} [\delta \tilde C']\; \langle K^{\ast 0} |\bar b
\gamma_\mu (1 + \gamma_5) s | B^0 \rangle \langle \phi | \bar s \gamma^\mu
s | 0 \rangle \ ,
\label{heff2}
\end{eqnarray}
where the factors in square brackets involve combinations of the
coefficients $\tilde C_i$, $\delta\tilde C_i$ and $\delta\tilde C'_i$ in
\eq{Heff} (explicit expressions, which are not important at this point,
will be given in the next Section). The information on high energy
contributions to the decay process ---including SM and NP interactions---
is carried by the Wilson coefficients, whereas low energy interactions are
taken into account in the hadronic matrix elements. From \eq{heff2}, it is
seen that the factorized matrix elements $\langle K^{\ast
0}|J_\mu|B^0\rangle$ can be separated into two terms, namely those driven
by vector and by axial-vector currents. Taking into account the Lorentz
decomposition in \eq{hlam}, the latter are found to contribute to
coefficients $a$ and $b$, and the former to $c$. In this way, one arrives
to the following simple relations between the various parameters in
\eq{coefficients}:
\begin{equation}
\frac{a_1}{c_1} \; = \; \frac{a_2}{c_2} \; = \; - \frac{a_3}{c_3} \ .
\label{relation}
\end{equation}
We recall that the decay $\bk$ is mainly driven by penguin-like operators.
Since $O_{11,12}$ and $O'_{11,12}$ do not contribute, the operators
corresponding to contributions 2 and 3 are nothing but the SM operators
and their parity-conjugated (i.e.\ those obtained by changing
$1\pm\gamma_5\to 1\mp\gamma_5$), respectively. Hence the sign difference
in \eq{relation} will allow in principle to distinguish between NP
contributions arising from effective left-handed (i.e.\ SM-like) and
right-handed currents.

Regarding Eqs.~(\ref{as}) and (\ref{obs}), it is clear that the above
relations between the contributions to $a$, $b$ and $c$ will become
particularly relevant for the triple products $\Lambda_{\bot j}$ and
$\Sigma_{\bot j}$, with $j=\|,0$. Let us focus first on $\Lambda_{\bot
\|}$ and $\Sigma_{\bot \|}$, and then comment on the other two. As stated,
with good approximation the process under study carries only one global
weak phase within the SM, therefore CP-odd observables will vanish in
the absence of NP. This is the case of $\Lambda_{\bot\|}$. From
\eq{coefficients}, using the relations in \eq{relation} one has \bea
\Lambda_{\bot\|} & = & -\, 8 \sqrt{x^2 -1} \; \bigg[\; a_1\, c_2\;\sin
(\delta_a - \delta_c) \;\sin (\tilde \delta_2 - \tilde \delta_1) \;
\sin (\phi_2 - \phi_1) \ + \nonumber \\
& & \hspace{2.4cm} a_1\, c_3\; \cos (\delta_a - \delta_c)\; \cos (\tilde
\delta_3 - \tilde \delta_1)\; \sin (\phi_3 - \phi_1) \ - \nonumber \\
& & \hspace{2.4cm} a_2\, c_3\; \cos(\delta_a - \delta_c)
\;\cos (\tilde \delta_2 - \tilde \delta_3)\; \sin (\phi_2-\phi_3) \bigg] \ .
\label{lam}
\eea
In the framework of factorization, one expects strong FSI phases to be
relatively small. However, notice that $\Lambda_{\bot\|}$ may be still
non-vanishing even in the limit where these are zero. This is a particular
feature of triple products, and an equivalent observable is not to be
found in $PP$ or $VP$ decay channels of $B$ mesons. As stated, another
relevant observable (in this case CP-even) is $\Sigma_{\bot \|}$, which
in our framework is given by
\begin{eqnarray}
\Sigma_{\bot\|} & = &
8 \sqrt{x^2 -1} \; \bigg[\; \frac12\; (a_1\, c_1\,+\,a_2\, c_2\,+\,a_3\, c_3)
\,\sin (\delta_a - \delta_c)\; + \nonumber \\
& & \hspace{2.1cm} a_1\, c_2\,\sin (\delta_a - \delta_c) \;\cos (\tilde \delta_2 - \tilde \delta_1) \, \cos (\phi_2 -\phi_1) \ - \nonumber \\
& & \hspace{2.1cm} a_1\, c_3\; \cos (\delta_a - \delta_c)\, \sin (\tilde\delta_3 - \tilde\delta_1) \, \cos(\phi_3 -\phi_1) \ + \nonumber \\
& & \hspace{2.1cm} a_2\, c_3\, \cos(\delta_a - \delta_c)\,
\sin (\tilde \delta_2 - \tilde \delta_3)\,\cos (\phi_2-\phi_3)\bigg]\; .
\label{sig}
\end{eqnarray}
Notice that this observable is different from zero only in the presence of
non-vanishing strong phases. In particular, within the SM one needs FSI
phases $\delta_a$ or $\delta_c$ to be nonzero ---and to be different from
each other---, therefore the value of $\Sigma_{\bot \|}$ is regarded as a
measure of FSI effects~\cite{Bab04,Bel04}.

As stated, within factorization it is natural to assume that strong FSI
phases are small, i.e.\ $\delta\ll 1$. Moreover, we will also assume
$\tilde\delta\ll 1$, since these phases arise in general from absorptive
contributions which are higher order in perturbation theory. Then from
\eq{lam} it can be seen that, depending on the nature of NP contributions,
one obtains sizable different behaviors in $\Lambda_{\bot\|}$. Indeed, the
contribution to $\Lambda_{\bot\|}$ arising from a combination of
coefficients of the same chirality [first term on the r.h.s.\ of \eq{lam}]
is suppressed by a factor $\sim\delta\tilde\delta$ with respect to the
remaining two (chirality-mixing) terms. In this sense, it could be said
that $\Lambda_{\bot\|}$ behaves as a filter of right-handed NP.
On the other hand, in the case of $\Sigma_{\bot \|}$ it is seen that all
contributions are of order $\delta$ or $\tilde\delta$. For this observable
it is also necessary to take into account the relative magnitude between
the various coefficients, the SM term (carrying the combination $a_1\,
c_1$) being the dominant one. As stated above, NP contributions are
expected to be suppressed by a factor $\xi\sim \delta C/C\sim \delta
C'/C$, which leads to $\Sigma_{\bot \|}\sim a_1\,c_1\, [ \delta + {\cal
O}(\delta\,\xi\,,\;\tilde\delta\,\xi)]$. Notice that the leading term in
the square brackets is independent of the nature of NP.

In order to make use of the chirality-filtering property of
$\Lambda_{\bot\|}$ when comparing with experiment, it is desirable to get
rid of the coefficients $a_i\,c_j$ which in general will be hardly
determined from the theory. In this way, we find convenient to consider
the ratio $r_\| \equiv \Lambda_{\bot \|} / \Sigma_{\bot \|}$, which has
two main advantageous features: $(i)$ it is sensitive to different NP
chiral structures, and depends on the coefficients through the ratios
$c_2/c_1\sim c_3/c_1\sim \xi$ [see Eqs.\ (\ref{r1a}-\ref{r1c})]; $(ii)$
since the order of magnitude of the denominator does not change with the
chirality of NP, it does not spoil the filtering properties of
$\Lambda_{\bot \|}$, and the ratio $r_\|$ becomes enhanced in the case of
right-handed NP currents. Indeed, one could have three different
situations upon the NP scenario considered:
\begin{eqnarray}
\mbox{SM} & \ \longrightarrow \ & r_\| \; \simeq \; 0 \label{r1a} \\
\mbox{only left-handed NP} &\longrightarrow & r_\| \, \simeq \,
     -2\; (\tilde \delta_2 - \tilde\delta_1) \; \frac{c_2}{c_1} \, \sin (\phi_2 - \phi_1)
     \, \sim \, \xi \; \tilde\delta \label{r1b}\\
     \mbox{right-handed\ NP} & \longrightarrow & r_\| \;
     \simeq \, -\frac{2}{\delta_a - \delta_c} \; \frac{c_3}{c_1}
     \, \sin(\phi_3 - \phi_1) \; \sim\; \xi\, /\, \delta \ .
\label{r1c}
\end{eqnarray}
Two main conclusions come out from these relations: first, a non-vanishing
measurable value of $r_\|$ represents a signature of NP. Second, this
observable is particularly sensitive to the presence of NP if nonstandard
interactions lead to effective right-handed quark currents. Notice the
interplay between the characteristic scale of NP and the strong phases
involved in the decay: if FSI phases $\delta$ are small, but still
measurable, the effect of right-handed currents should be observable when
$\xi \gtrsim \delta$. In contrast, SM-like contributions to $r_\|$ arising
{}from NP at this same scale would be additionally suppressed by strong
phases originated in the high energy region. In this way, the possible
detection of right-handed NP through this observable is intimately
associated with the relation between the NP scale and the FSI phases, $\xi
\leftrightarrow \delta$. We notice at this point that a comparison of the
predictions for $r_\|$ with experimental data is not possible yet, since
present measurements~\cite{Bab04,Bel04} of $\Sigma_{\bot\|}$ give values
which are still compatible with zero. However, given the potential power
of $B$ factories, we expect that the experimental analyses will be able to
set bounds for $r_\|$, providing a clue about the nature of NP.

Let us now comment on the other two triple products, $\Lambda_{\bot 0}$
and $\Sigma_{\bot 0}$, and also the ratio $r_0 \equiv \Lambda_{\bot
0}/\Sigma_{\bot 0}$. Regarding the definitions in
Eqs.~(\ref{as}-\ref{obs}), it is seen that a similar behavior to their
cousins $\Lambda_{\bot\|}$, $\Sigma_{\bot\|}$ should be expected, even if
the results will be more involved since the amplitude $A_0$ receives
contributions from both Lorentz invariant parameters $a$ and $b$. In fact,
it is easy to see that $\Lambda_{\bot 0}$ acts as a filter of right-handed
NP currents, in a similar way as $\Lambda_{\bot\|}$, in the limit of small
strong phases. The ratio $r_0$, upon the nature of the NP, behaves as
follows:
    \bea
    \mbox{SM} &\longrightarrow & r_0 \  \simeq \ 0 \label{r2a}\\
    \mbox{only left-handed NP} &\longrightarrow & r_0 \ \simeq \
     -2\; (\tilde \delta_2 -\tilde\delta_1) \; \frac{c_2}{c_1} \, \sin (\phi_2 - \phi_1)
     \ \sim \ \xi \; \tilde\delta \label{r2b} \\
    \mbox{right-handed NP} & \longrightarrow & r_0 \ \simeq \
    \frac{2[x a_1 + (x^2-1) b_1]}{x(\delta_c - \delta_a) a_1 +
    (x^2-1) (\delta_c - \delta_b) b_1} \, \frac{c_3}{c_1} \sin (\phi_3 - \phi_1)
    \nonumber \\
    & & \quad\ \, \sim \ \xi\, / \, \delta
    \label{r2c}
    \eea Thus, the same qualitative behaviour found for $r_\|$,
c.f.~Eqs.~(\ref{r1a}-\ref{r1c}), holds for $r_0$ as well. Notice, however,
that the combinations of the parameters $a_1$ and $b_1$ and the FSI phases
in \eq{r2c} may lead to cancellations which modify the naively expected
orders of magnitude. Thus, in this sense, $r_0$ (which represents a better
observable from the experimental point of view) will be not as conclusive
as $r_\|$. We recall that present experiments~\cite{Bab04,Bel04} have
measured a nonzero value for $\Sigma_{\bot 0}$, allowing to establish
first experimental bounds on the value of $r_0$, namely $r_0=-0.14\pm
0.21$~\cite{Bel04}\footnote{Notice that the experimental values reported
in Ref.~\cite{Bel04} result from the average between the data obtained
from $\bk$ and $B^\pm \to \phi K^{\ast\pm}$, which is justified if the
annihilation contribution to $B^\pm \to \phi K^{\ast\pm}$ can be safely
neglected. If this is not the case~\cite{Epe03}, one should take into
account for our purposes only the data corresponding to the neutral
channel.}.

To conclude this section, some final remarks are in order. First, notice
that the chirality-mixing suppression in the $\Lambda_{\bot j}$'s could be
modified by the effect of large nonfactorizable contributions, which have
been neglected in our analysis. The consistency of the factorization
approximation can be studied by performing a global analysis of the
various observables that can be measured in this decay channel, as well as
in related processes such as $B^\pm\to\phi K^{\ast\pm}$~\cite{Epe03}.
Second, we point out that a key feature of the neutral channel analyzed
here is that it has a unique way to factorize the hadronic currents in the
factorization approximation, and this fact makes possible the chirality
test. As a matter of fact, this same analysis could be applied to the
channels $B^0\to K^{\ast 0}\bar K^{\ast 0}$, $B^0\to\phi\phi$,
$B^0\to\rho^0\phi$ and $B^0\to\omega\phi$, which have not been
experimentally observed yet. Finally, as mentioned in Sect.~II, we stress
that the existence of other left- or right-handed NP contributions may be
still worked out within this framework leaving our conclusions unchanged,
since the property of filtering right-handed currents relies on the
structure of the observables $\Lambda_{\bot j}$ and does not depend on the
number of contributions considered.

\section{Numerical estimations in a LR symmetric model}

Theories based upon the electroweak gauge group $SU(2)_L \times SU(2)_R
\times U(1)$ represent well-known extensions of the Standard Model. We
analyze here the so-called Left-Right Symmetric Model (LRSM)~\cite{lrsm},
in which the elements of left and right quark mixing matrices are equal in
modulus, i.e.\ $|V^L_{ij}| = |V^R_{ij}|$. This model has been largely
discussed in connection with penguin-dominated decays $\bar b\to \bar s q
\bar q$~\cite{works}, and serves as a good example to illustrate the
special features of the observables $\Lambda_{\bot\|}$ and
$\Sigma_{\bot\|}$ (and their ratio $r_\|$) considered in the preceding
Sections.

The effective Hamiltonian for $\bar b\to \bar s q \bar q$ decays within
the LRSM has been calculated in LL precision by Cho and
Misiak~\cite{Cho93}. Keeping only the top and bottom quark masses as
non-vanishing, the matching conditions at the $W$ scale lead to
        \be
        \begin{array}{rclcrcl}
        C_2 (M_1)&=&1\;, & &C'_2 (M_1)&=&0\;, \\
        C_\gamma (M_1)&=&D'_0 (x) + A_{tb}^\ast\, \tilde D'_0(x)\;,
        & &C'_\gamma (M_1)&=& A_{ts}\, \tilde D'_0 (x)\;,\\
        C_g (M_1) &=& E'_0 (x) + A_{tb}^\ast\, \tilde E'_0 (x)\;,
        & &C'_g (M_1)
        &=& A_{ts}\, \tilde E'_0 (x)\;,
        \end{array}
        \ee
whereas the remaining coefficients $C_i$ and $C'_i$ are equal to zero.
Here $M_1$ is the mass of the $W_1^\pm$ gauge bosons (equivalent to the
standard $W^\pm$), $E'_0(x)$ and $D'_0(x)$ are SM Inami-Lim functions, and
their left-right analogues are denoted by $\tilde E'_0 (x)$ and $\tilde
D'_0 (x)$ (see Ref.~\cite{Cho93}). These functions are evaluated at
$x=m_t^2/M_1^2$ , while the coefficients $A_{tb}$ and $A_{ts}$ are given
in terms of LRSM parameters,
        \bea
        A_{tb}=\xi \frac{m_t}{m_b}\frac{V_{tb}^R}{V_{tb}^L}
        e^{i\omega} \equiv \xi \frac{m_t}{m_b}e^{i\sigma_1} \\
        A_{ts}=\xi \frac{m_t}{m_b}\frac{V_{ts}^R}{V_{ts}^L}
        e^{i\omega} \equiv \xi \frac{m_t}{m_b}e^{i\sigma_2}\ .
        \eea
Here $\xi$ is the $W_L$-$W_R$ mixing angle times the ratio between right
and left coupling constants, while $\sigma_{1,2}$ are unknown CP-violating
phases that take values in the range $[0,2\pi ]$ (for simplicity we take
the SM amplitude to be real). The mass of the gauge bosons $W_2^\pm$ does
not appear here explicitly since its lower bound, in the multi-hundred GeV
region, leads to negligible contributions to the effective Hamiltonian.

The renormalization group mixing is governed by a $20\times 20$ anomalous
dimension matrix $\gamma$, which decomposes into two identical $10\times
10$ submatrices. The expression for the SM $8\times 8$ submatrix can be
found in Ref.~\cite{ref30} and the rest of the entries have been computed
in Ref.~\cite{Cho93}. The running of the Wilson coefficients to the $m_b$
scale in the LL approximation for five active flavors yields
\bea
C_i(\mu=m_b) = \sum_{k,l} (S^{-1})_{ik} \eta^{3\lambda_k/4}
S_{kl} C_l (M_1)\ ,
\end{eqnarray}
where the $\lambda_k$'s in the exponent of $\eta =
\alpha_s(M_1)/\alpha_2(m_b)$ are the eigenvalues of the anomalous
dimension matrix, and $S$ is a matrix containing the corresponding
eigenvectors. The explicit computation of these Wilson coefficients at the
$m_b$ scale has been performed in Ref.~\cite{barenboim2}.

In order to evaluate the relevant hadronic matrix elements in the case of
the decay $\bk$, once again we make use of the factorization
approximation. Then, using the equations of motion, it is seen that the
matrix elements of magnetic dipole operators can be written in terms of
the penguin ones as
        \bea
        \langle O^G_8 \rangle &=& -\frac{\alpha_s}{4\pi}
        \frac{m_b}{\sqrt{\langle q^2\rangle}}
        \left[ \langle O_4 \rangle + \langle O_6 \rangle -
        \frac{1}{N_c} ( \langle O_3 \rangle + \langle O_5 \rangle ) \right], \\
        \langle O^\gamma_7 \rangle &=& -\frac{\alpha}{3\pi}
        \frac{m_b}{\sqrt{\langle q^2\rangle}}
        \left[ \langle O_7 \rangle + \langle O_9 \rangle  \right]\ ,
        \label{dipole}
        \eea
and similar equations hold for their primed counterparts. In this way, one
can get rid of the magnetic dipole operators when writing the $\bk$ decay
amplitude by including their contributions into the Wilson coefficients
$C^{\rm eff}_i$ and ${C'}^{\rm eff}_i$, with $i=3,\dots ,10$. The
amplitude can be written as~\cite{Ali98}
        \bea
        A(B^0\to\phi K^{\ast 0}) &=& \langle H_{eff} \rangle \nonumber \\
        &=& -\,\frac{G_F}{\sqrt 2}\; V_{ts}^L V_{tb}^{L\ast}
        \left[ a_3 + a_4 + a_5 -\frac{1}{2} (a_7 + a_9 + a_{10}) \right]
        X_L^{(B K^\ast,\phi)} \nonumber \\
        && -\frac{G_F}{\sqrt 2}\; V_{ts}^L V_{tb}^{L\ast} \left[ a'_3 + a'_4
        + a'_5 -\frac{1}{2} (a'_7 + a'_9 + a'_{10}) \right] X_R^{(B K^\ast,\phi)}\ ,
        \label{perales} \eea
where $X_{L,R}^{(B K^\ast,\phi)} = \langle K^{\ast 0} | \bar b \gamma^\mu (1\mp
\gamma_5) s |B^0\rangle \langle \phi | \bar s \gamma_\mu s | 0 \rangle\,$,
        \bea
        a_{2i-1} = C^{\rm eff}_{2i-1} + \frac{1}{N_c} C^{\rm eff}_{2i},
        \ \ \ \ a_{2i} = C^{\rm eff}_{2i} + \frac{1}{N_c} C^{\rm
        eff}_{2i-1}\ ,
        \label{aes}
        \eea
and similar relations are satisfied by the coefficients $a'_i$. Taking now
$\sqrt{\langle q^2 \rangle} =m_b/\sqrt 2$ in
Eqs.~(\ref{dipole})~\cite{Ali98}, together with $m_t/m_b = 60$ and $N_C =
3$, the amplitude reads
        \bea
        A(B^0\to\phi K^{\ast 0}) & \simeq & -\,\frac{G_F}{\sqrt 2}\,
        V_{ts}^L V_{tb}^{L\ast} \left( -0.019\; +\; 0.12
        \; \xi\;  e^{-i\sigma_1} \right) X_L^{(B K^*,\phi)} \nonumber \\
        && -\,\frac{G_F}{\sqrt 2}\, V_{ts}^L V_{tb}^{L\ast}
        \left( 0.12\; \xi\; e^{i\sigma_2} \right) X_R^{(B K^*,\phi)} \ .
        \label{serrat}
        \eea
Notice that, aside from the scale suppression $\xi$, the NP contributions
turn out to be enhanced. This is mainly given by a factor $m_t/m_b$, which
arises in the case of $V+A$ interactions since the usual helicity flip is
not needed in penguin amplitudes.

In order to extract from \eq{serrat} the different coefficients and phases
in \eq{coefficients}, we write the factorized matrix elements in terms of
form factors $f_V$, $V^{B\to V}(q^2)$, and $A_i^{B\to V}(q^2)$, $i =
0,1,2$:
        \begin{eqnarray}
        \langle V (\varepsilon,p') | V_\mu | 0 \rangle & \!=\! & f_V \, m_V \,
        \varepsilon^\ast_\mu \nonumber \\
        \langle V (\varepsilon, p') | V_\mu | B (p) \rangle & \!=\! &
        -\,\frac{2}{m_V + m_B} \; \epsilon_{\mu\nu\alpha\beta}
        \, \varepsilon^{\ast\,\nu} \,
        p^\alpha {p'}^\beta \, V^{B\to V}(q^2) \nonumber \\
        \langle V (\varepsilon, p') | A_\mu | B (p) \rangle & \!=\! & i\,
        \frac{2 m_V (\varepsilon^\ast\cdot q)}{q^2}\;
        q_\mu\, A_0^{B\to V}(q^2) +
        i\, (m_V + m_B) \left[\varepsilon^\ast_\mu - \frac{(\varepsilon^\ast
        \cdot q)}{q^2}\, q_\mu \right] \! A_1^{B\to V} (q^2) \nonumber \\
        & & -\, i \left[(p + p')_\mu\, - \frac{(m_B^2-m_V^2)}{q^2}
        \;q_\mu\right] \frac{(\varepsilon^\ast\cdot q)}{m_V + m_B} \;
        A_2^{B\to V}(q^2)\ .
        \label{ale}
        \end{eqnarray}
Here $V(\varepsilon,p')$ stands for the outgoing vector mesons $\phi$ or
$K^{\ast 0}$, $V_\mu$ and $A_\mu$ are the corresponding vector and
axial-vector quark currents and $q=p-p'$ is the transferred momentum. The
vector and axial-vector form factors can be estimated from the analysis of
semi-leptonic $B$ decays, using the ansatz of pole dominance to account
for the momentum dependencies in the region of interest. Taking into
account Eqs.~(\ref{serrat}) and (\ref{ale}) and comparing with
Eqs.~(\ref{hlam}) and (\ref{coefficients}) one can obtain explicit
expressions for the coefficients $a_i$ and $c_i$, needed to estimate the
values of the observables $\Lambda_{\bot\|}$ and $\Sigma_{\bot\|}$. We
find
    \bea
        a_1 &=& - 0.019\; \kappa \; ( m_B + m_{K^\ast} ) \;
        A_1^{B\to K^\ast}(m_\phi^2) \ , \nonumber \\
        c_1 &=& 0.019 \; \kappa \;
        \left(\frac{2\, m_{K^\ast}\, m_\phi}{m_B + m_{K^\ast}} \right)
        \; V^{B\to K^\ast}(m_\phi^2) \ ,\nonumber \\
        a_2 &=&  0.12 \; \xi \; \kappa \; ( m_B + m_{K^\ast} )
        \; A_1^{B\to K^\ast}(m_\phi^2) \ ,\nonumber \\
        c_2 &=& -0.12 \; \xi \; \kappa \;
        \left(\frac{2\, m_{K^\ast}\, m_\phi}{m_B + m_{K^\ast}} \right)
        \; V^{B\to K^\ast}(m_\phi^2) \ ,\nonumber \\
        a_3 &=& - a_2 \ ,\nonumber \\
        c_3 &=& c_2 \label{c3} \ ,
\end{eqnarray}
where $\kappa = i \frac{G_F}{\sqrt{2}} V_{ts}^L V_{tb}^{L\ast} m_\phi
f_\phi\,$. The weak phases in \eq{coefficients} are given by
$\phi_2=-\sigma_1$ and $\phi_3=\sigma_2$, the absorptive phases $\tilde
\delta_i$ have been neglected in this approximation, and the strong FSI
phases $\delta$, which are expected to be small, are left as unknown
parameters. As expected, the coefficients in Eqs.~(\ref{c3})
satisfy the relations in \eq{relation}. Finally, the observables
$\Lambda_{\bot\|}$ and $\Sigma_{\bot\|}$ can be estimated in the LRSM by
making use of Eqs.~(\ref{lam}) and (\ref{sig}). This leads to
\bea
\hspace{-.9cm} \Lambda_{\bot\|} & = &  -\, 16 \sqrt{x^2-1}\; \kappa^2
\,m_{K^\ast}\, m_\phi\, V^{B\to K^\ast}(m_\phi^2)\, A_1^{B\to
K^\ast}(m_\phi^2) \times
\nonumber \\
& & \hspace{4cm} \left[ 0.019 \; \times \; 0.12\; \xi\; \sin\sigma_2 \; -
\; (0.12)^2\; \xi^2 \; \sin (\sigma_1 + \sigma_2) \right]\; ,
\label{kmenos} \\
\hspace{-.9cm} \Sigma_{\bot\|} & = &  8 \sqrt{x^2-1}\; \kappa^2 m_{K^\ast} \, m_\phi
\, V^{B\to K^\ast}(m_\phi^2)\, A_1^{B\to K^\ast}(m_\phi^2) \,  (0.019)^2 \;
(\delta_c - \delta_a)\ + \ {\cal O} (\xi \delta )\ .
\label{kmas} \eea
\eq{kmenos} deserves some discussion. As it can be easily seen, the
dependence of $\Lambda_{\bot\|}$ on $\sigma_1$ is suppressed by a factor
$\xi$ with respect to that on $\sigma_2$. In fact, this could be expected
since the right-handed NP is incorporated in the effective Hamiltonian
through the dipolar operators, which are obtained by a mass insertion in
the $b$ quark line, whereas the contribution of the $s$ quark is
neglected. Therefore, although the SM-like NP operators and the
right-handed ones are both proportional to $\xi$, the former depend only
on $\sigma_1$ and the latter only on $\sigma_2$. Since the structure of
$\Lambda_{\bot\|}$ is such that suppresses in $\delta\tilde\delta(=0)$ the
mixture of terms having the same chirality, the only combined
contributions that remain are those given by SM $\times$ Right-NP ($\sim
\xi$) and Left-NP $\times$ Right-NP ($\sim \xi^2$).

Using Eqs$.\,$(\ref{kmenos}-\ref{kmas}) we obtain the value for
$r_\|=\Lambda_{\bot\|} / \Sigma_{\bot\|}$ within the LRSM to leading order
in $\xi$,
\begin{equation}
r_\|=\frac{0.12}{0.019}\; \frac{2 \xi}{\delta_a-\delta_c}
\sin \sigma_2\ .
\label{tanto}
\end{equation}
Notice that in the ratio $r_\|$ one gets rid of form factors, which are in
general theoretically uncertain. In \eq{tanto}, $\sigma_2$ is a free
parameter, hence it is natural to assume $\sin\sigma_2\simeq {\cal O}(1)$.
The scale of NP is given basically by the mass of the gauge bosons
$W_2^\pm$; present limits lead to an upper bound for $\xi$ of about
$0.04$~\cite{pdg}. Clearly, the value of $r_\|$ turns out to be enhanced
by the FSI phases (assumed to be small) in the denominator. As discussed
in the previous section, in order to distinguish right-NP from left-NP by
means of $r_\|$, it is essential to consider the relation between the NP
scale and the FSI phases $\delta$.

As mentioned, present data are not precise enough to place experimental
bounds on $r_\|$. However, in view of the positive result in the case of
$\Sigma_{\bot 0}$, it is natural to expect that future measurements
provide a nonzero value for the combination $\delta_a - \delta_c$, and
hence a nonzero $\Sigma_{\bot\|}$. We expect that forthcoming experimental
data will be able to place bounds on $r_\|$, and in this way quest for the
existence of right-handed NP currents as those proposed by the LRSM.

\section{Conclusions}

In this work we have studied possible signatures of New Physics through
the angular analysis of the decay $B^0 \to \phi K^{*0}$, which is shown to
be an optimal channel to search for nonstandard effects. In particular, we
have studied how the effects of left and/or right-handed NP currents show
up in some selected observables.

In the construction of the effective Hamiltonian we have considered,
besides the SM physics, possible NP contributions to SM-like operators, as
well as contributions driven by new operators arising from the presence of
right-handed NP currents. We have worked within the framework of
factorization, considering for simplicity the case in which one has only
one NP contribution of each chirality. Since the decay under study has
only one possible way to factorize, the nonstandard contributions are
subject to kinematic cancellations in some observables, allowing to obtain
a chirality filter for the NP.

The coefficients in the angular distributions of the decay of a $B$ into
two vector mesons, such as $\bk$, give a rich variety of observables. We
have found that if both the absorptive and FSI strong phases ($\tilde
\delta$ and $\delta$, respectively) as well as the ratio between SM and NP
scales ($\xi$) are assumed to be small quantities, then one can define
ratios which show a strongly different behavior upon the nature of the NP.
In particular, we have concentrated on the T-odd triple products
$\Lambda_{\bot i}\ (i=0,\|)$, which can be written as sums of various
terms including contributions from different invariant amplitudes. The key
is to realize that those terms that contain a mixture of two contributions
coming from operators with the $same$ chiral structure are suppressed by
an order of magnitude of $\sim\,\delta\,\tilde\delta$ with respect to
those which arise from the mixture of operators with {\em different}
chiralities. In this way, they effectively serve as filters of
right-handed NP. We have also considered the CP-conserving observables
$\Sigma_{\bot i}$, which at leading order behave as $\sim\,\delta$
independently of the nature of the NP. As explained in the text, we
propose to measure the ratio $r_i = \Lambda_{\bot i} / \Sigma_{\bot i}$,
which shows a different behavior upon the nature of the NP, namely \bea
r_i \sim \left\{
\begin{array}{lcl}
0&& \mbox{SM} \\
\xi \tilde \delta && \mbox{only Left-NP} \\
\xi/\delta && \mbox{Right-NP}
\end{array}
\right. \label{cuco} \eea Although $r_i$ should be handled with care,
since it is a ratio between two small quantities, from \eq{cuco} it is
seen that this represents an optimal observable for the search of effects
of right-handed NP, provided that strong FSI phases $\delta$ are found to
be nonzero but sufficiently small. On the other hand, if NP effects are
observed in any other process, a measurement of a tiny upper bound for $r_i$
should be taken as a strong indication in favor that the underlying theory
leads to effective operators of the same type of the SM. In addition,
\eq{cuco} shows that the perceptiveness of right-handed NP is strongly
dependent on the relation between the relative scale of NP and the strong
FSI phases. We notice that although the result in \eq{cuco} is clean for
the ratio $r_\|$, it should be taken with some care for $r_0$, since a
fine tuning of the parameters could lead to cancellations modifying the
naively expected orders of magnitude. On the experimental side, present
measurements of both $\Sigma_{\bot\|}$ and $\Lambda_{\bot\|}$ are still
compatible with zero, and only a positive result has been achieved in the
case of $\Sigma_{\bot 0}$. We expect forthcoming experiments to establish
precise bounds for these observables in the near future, providing clues
about the nature of new physics beyond the SM. Our analysis has been
complemented with a numerical estimation of the ratio $r_\|$ in a LRSM,
which shows that the allowed parameter space for these models is
compatible with the presence of large NP effects in penguin-dominated
modes like $\bk$.

\section*{Acknowledgments}

We thank A. Gritsan for helpful information on experimental results, A.\
Kagan and D.\ London for useful comments on the paper, and J.\ Bernab\'eu
for a revision of the manuscript. A.E.\ acknowledges the warm hospitality
offered by the Physics Department of the University of La Plata,
Argentina. This work has been partially supported by CONICET and ANPCyT,
under grants PIP 2873 and PICT 03-10718/02, respectively. Financial aid
has also been received from Fundaci\'on KONEX (A.E.) and Fundaci\'on
Antorchas (A.E.\ and A.S.), Argentina.


\end{document}